\documentclass[aps,twocolumn,nofootinbib,amsmath,amssymb,prl,showpacs,superscriptaddress]{revtex4-1}
\usepackage{graphicx,graphics}
\usepackage{bm}
\usepackage{amssymb}   
\usepackage{xcolor, colortbl}
\usepackage{textcomp} 
\usepackage{natbib}
\usepackage{array}
\usepackage{verbatim}
\usepackage{dcolumn}
\usepackage{amsmath,amssymb,amsfonts}
\usepackage{latexsym,verbatim}
\usepackage{bm}
\usepackage{color}
\usepackage{ulem}
\usepackage{graphicx,epstopdf} 
\usepackage[breaklinks=true,colorlinks,citecolor=blue,linkcolor=blue,urlcolor=blue]{hyperref}
\definecolor{darkseagreen1}{rgb}{0.82,0.94,0.82}

\begin{document}

\title{Strongly bound Mott-Wannier Excitons in GeS and GeSe monolayers.}

\author{Lidia C. Gomes}
\affiliation{Graphene Research Centre and CA2DM, National University of Singapore, 117542, Singapore}

\author{P. E. Trevisanutto}
\affiliation{Graphene Research Centre and CA2DM, National University of Singapore, 117542, Singapore}
\affiliation{Singapore Synchrotron Light Source, National University of Singapore, 5 Research Link, Singapore 117603, Singapore}

\author{A. Carvalho}
\affiliation{Graphene Research Centre and CA2DM, National University of Singapore, 117542, Singapore}
\author{A. S. Rodin}
\affiliation{Graphene Research Centre and CA2DM, National University of Singapore, 117542, Singapore}
\author{A. H. Castro Neto}
\affiliation{Graphene Research Centre and CA2DM, National University of Singapore, 117542, Singapore}

\date{\today}

\begin{abstract}
The excitonic spectra of single layer GeS and GeSe are predicted by  {\it ab initio} GW-Bethe Salpeter equation calculations.
$G_0W_0$ calculations for the band structures find a fundamental band gap of 2.85 eV for GeS and 1.70 eV for GeSe monolayer.
However, excitons are tightly bound, specially in GeS at the $\Gamma$ point, where the quasi-particle interactions
are so strong that they shift the  $\Gamma$ exciton peak energy into the visible range and below
the off-$\Gamma$ exciton peak.
The lowest energy excitons in both materials are excited by light along the zigzag direction and have exciton
binding energies of  1.05 eV and 0.4 eV, respectively,
but despite the strong binding, the calculated binding energies are in agreement with a Mott-Wannier model.

\end{abstract}


\maketitle


\hyphenation{ALPGEN}
\hyphenation{EVTGEN}
\hyphenation{PYTHIA}

{\it Introduction--}
The excitonic properties of 2D semiconductors have revealed to be fundamentally different
from bulk semiconductors.
In the 2D limit, the electron and hole experience reduced Coulomb screening
as the dielectric environment changes abruptly from the layer to the vacuum.~\cite{Keldysh1979cii,Cudazzo2011dsi,Berkelbach2013ton,Castellanos_Gomez2014iac}
Additionally, the confinement of the electron and hole
to the plane also contributes to increasing the exciton binding energy. 
As a result, the optical properties of 2D semiconductors such as 
transition metal dichalcogenides, phosphorene and group-IV monochalcogenides are
dominated by excitonic effects.
Thus, the use of 2D materials for optoelectronic applications requires a deeper understanding
of the excitonic properties.

A few experimental studies to date have measured the exciton binding
energies of monolayer MoS$_2$, WS$_2$, MoSe$_2$ and WSe$_2$ 
to be in the range of 0.2-0.8 eV\cite{zhang-NL-14-2443,ye-N-513-214,ugeda-NM-13-1091,wang-PRL-114-097403,zhu-SR-5-9218}, 
even though the values for isolated monolayer are expected, according to theory, to be around 1 eV.\cite{komsa-PRB-86-241201,louie}
Another outcome of the non-local dielectric screening
is that the energy level ordering for the exciton 1s, 2s, 2p, etc. states differs from a Hydrogenic Rydberg series.\cite{chernikov-PRL-113-076802,ye-N-513-214}
In nearly neutral monolayer samples of semiconducting transition metal dichalcogenides,
other quasiparticles (QP) have been observed, including negative and positively charged trions\cite{shang,mak-NM-12-207,zhu-SR-5-9218}
and biexcitons.\cite{shang}
Such abundance of excitonic effects has no parallel in 3D systems.

In phosphorene, excitons were also found to be strongly bound,
with an exciton binding energy of 0.8-0.9 eV.\cite{PhysRevB.89.235319,PhysRevB.90.075429}
But different from transition metal dichalcogenides, phosphorene has in-plane anisotropy,
resulting in nearly unidimensional exciton wavefunctions,\cite{wang2015highly}
such that light emitted upon recombination of the lowest energy exciton
is linearly polarized along the light effective mass direction.

Group-IV monochalcogenides assume a structure similar to that of black phosphorus,
and therefore marked anisotropy of the optical properties is also expected.
However, even though determining bandgap values, optical absorption thresholds and identifying bound excitons is essential 
for the design of optoelectronic devices, determination of the number of layers,\cite{brent-JACS-137-12689}
and optical detection of the ferroelectric and ferroelastic state \cite{wu-NL-16-3236,wang2016two,hanakata2016memory},
 the optical properties of group-IV monochalcogenide monolayers are still object of discussion\cite{PhysRevB.92.085406, tritsaris-JAP-113-233507}.

In this letter, we predict both the GW quasi-particle band structures  and the absorption spectra of 
the group-IV monochalcogenide monolayers GeS and GeSe from first-principles,
highlighting the large exciton binding energy of GeS. In order to calculate the two-body electron-hole ($e$-$h$) Green's function, we have utilized the {\it ab initio} Many Body Perturbation Theory approach, the Bethe-Salpeter Equation (BSE) on top of GW self energy corrections (GW-BSE) \cite{rmp_onida}. In addition, we have analyzed the suitability of the Hyde, Scuseria and Ernzerhof (HSE) hybrid density functional\cite{HSE}  comparing the DFT and GW bandstructures. Our results display the presence of bound and localized excitons in both GeS and GeSe monolayer (with the binding exciton energy $~$ 1 eV and 0.4 eV, respectively). The Mott-Wannier model \cite{PhysRevB.90.075429} binding energies are in agreement with the GW-BSE. Moreover, we have determined the exciton binding energy trends for both GeS and GeSe in the presence of substrate dielectric constants.

{\it Computational details--}\label{Comput_details}
We use first-principles calculations to obtain the optimized structure and electronic bands of GeS and GeSe monolayers. A first-principles approach is employed based on Kohn-Sham density functional theory (KS-DFT)\cite{PhysRev.140.A1133}, as implemented in the {\sc Quantum ESPRESSO} code\cite{Giannozzi2009}. The exchange correlation energy was described by the generalized gradient approximation (GGA) using the PBE\cite{PhysRevLett.77.3865} functional, and the interactions between valence and core electrons are described by the  Troullier-Martins pseudopotentials\cite{PhysRevB.43.1993}. The Kohn-Sham (KS) orbitals are expanded in a plane-wave basis with a cutoff energy of 70~Ry. The Brillouin-zone (BZ) is sampled using a $\Gamma$-centered 10$\times$10$\times$1 grid, following the scheme proposed by Monkhorst-Pack (MP) \cite{PhysRevB.13.5188}. 
Structural optimization has been performed with a very stringent tolerance of 0.001~eV/\AA{}. 
In parallel, the HSE hybrid exchange-correlation functional\cite{HSE} is used to estimate the energy band gap, which is well known to be underestimated by standard DFT exchange-correlation functionals, including the generalized gradient approximations. 

The supercells are periodic in the monolayer plane and large vacuum regions ($>$ 10~\AA{}) are included to impose periodic boundary conditions in the perpendicular direction. Convergence tests were performed for the vacuum thickness, and the values used are enough to avoid spurious interaction between neighboring images.
 
 
Subsequently, the KS one-electron energies are corrected with the G$_0$W$_0$  ({\it one-shot}) self energy corrections $\Sigma$. These calculations are performed on top of DFT-PBE ground state ones as implemented in {\sc BerkeleyGW} code \cite{Deslippe20121269}. The BZ is sampled with a $32\times 32\times 1$ MP k-point mesh grid. The convergence is achieved with $300$ unoccupied states with a slab plane ($xy$) truncation of the bare Coulomb potential. 

The absorption spectrum is calculated as the imaginary part of the macroscopic dielectric function $\epsilon_{M}(\omega)$. Starting from the $GW$-$\Sigma$ corrections, the electron-hole (e-h) interactions are then included by using the Bethe Salpeter Equation (BSE) for the two particle correlation function $L$. Our GW-BSE calculations are restricted to the Tamm-Dancoff approximation which provides good results for semiconductors. The BSE kernel is evaluated first on a coarse k-grid ($32\times32\times1$) and then interpolated onto a finer grid ($64\times64\times1$).   


{\it Bandstructures--}\label{Bandstructure}
The electronic bandstructure of bulk GeS and GeSe have been discussed in previous theoretical works, where {\it ab initio} calculations indicate (underestimated) indirect gaps of 1.2~eV and 0.6~eV, respectively, at the DFT-GGA level~\cite{PhysRevB.92.085406}. More accurate results are achieved with HSE hybrid DFT functional and G$_{0}$W$_{0}$ approximation, and the corrected gaps agree well with available experimental data. For GeS, for example, theory indicates gaps between 1.53-1.81~eV, in close agreement with experimental values, in the range 1.70-1.96~eV for the conduction gap after extrapolation to $T=0$~\cite{PhysRevB.16.1616,kyriakos-SST-4-365,PhysRevB.87.245312}. The spread in experimental values for conduction and optical gaps makes an estimate of the exciton binding energy difficult to obtain, but place a higher bound at 0.3 eV.

\begin{figure}[!htb]
  \includegraphics[scale=0.40]{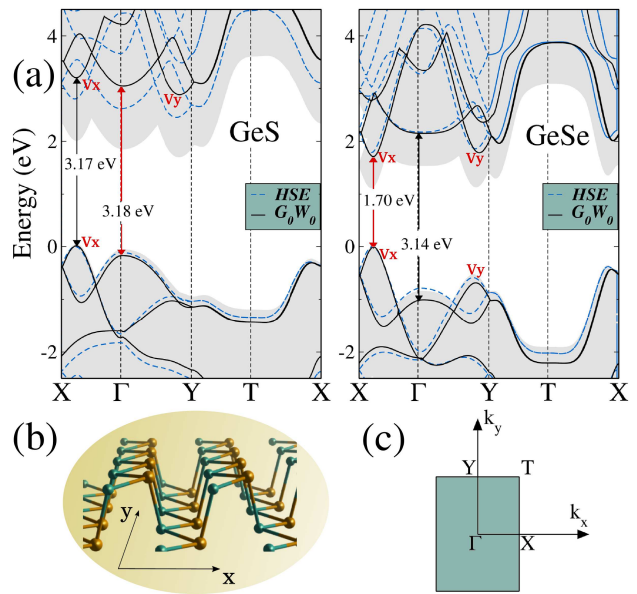}
   \caption{\small (Color online) (a) Electronic structure of GeS and GeSe monolayers calculated by GGA (full gray background), HSE (dashed blue lines) and GW (full black lines) methods. The direct transitions shown by the red arrows give rise to the excitons of highest binding energies. (b) The structure of the monolayers showing the armchair and zig-zag directions placed along the $x$ and $y$-axis, respectively. (c) Brillouin Zone with the high symmetry points.}
   \label{bands}
 \end{figure}

Single-layer GeS preserves the indirect gap character (Fig.~\ref{bands}), with the conduction band minimum localized at a point $V_y$ along the $\Gamma$-$Y$ line of the Brillouin zone (BZ) and the valence band maximum (VBM) at a point $V_x$ along $\Gamma$-$X$, the later very close in energy to a second maximum at $\Gamma$. Monolayer GeSe changes from indirect to direct gap along $\Gamma$-$X$. As discussed in Ref.~\onlinecite{PhysRevB.92.085406}, the corrections introduced by the hybrid functional result just in an increase of the gap energy, given by a rigid shift of the bands, while their dispersion are preserved. The minimum indirect and direct gaps of 2.45~eV and 1.79~eV are calculated at the HSE level for GeS and GeSe, respectively. 

In Fig.~\ref{bands}, the G$_{0}$W$_{0}$ corrections to the band structure are shown. Similar to the HSE results, the conduction and valence bands are rigidly shifted away from each other, with the minima and maxima located at the same positions in the high symmetry lines of the BZ as the DFT-GGA and HSE methods. The resulting indirect energy gap of 2.85~eV and direct gap of 1.70~eV for GeS and GeSe, respectively, show differences of 0.4~eV and 0.09~eV from the predicted values by HSE approximation to the exchange-correlation energy with an overall good agreement in the band values and band dispersions. 
In GeS, most importantly, the direct band gap values at $\Gamma$ and $V_x$ point are  nearly identical for all of three methods.

\begin{figure}[!htb]
    \centerline{
    \includegraphics[scale=0.30]{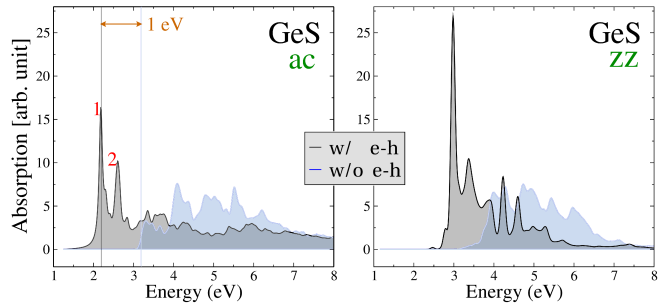}}
   \caption{\small (Color online) Absorption spectra of GeS monolayer with (gray spectrum) and without (blue spectrum) electron-hole interactions for light polarized along zigzag (zz) and armchair (ac) in-plane directions. Two exciton states (peaks 1 and 2) are formed inside the $G_0W_0$ gap along $x$. Peak 1 arises from direct transitions at $\Gamma$, while peak 2 are due to direct transitions at the $V_x$ valleys (along the $\Gamma$-X direction).}
   \label{abs_GeS}   
\end{figure}

{\it Absorption spectra--}\label{Abs_spect}
In Fig.~\ref{abs_GeS}, we present the absorption spectra of single-layer GeS calculated in GW-Random Phase Approximation (GW-RPA) (labelled ``without e-h interactions") and GW-BSE approach (``with e-h interactions"). 
Due to anisotropy, the absorption spectrum along $x$ (armchair direction) considerably differs from the polarization along $y$ (zig-zag direction).
In armchair direction, when the e-h interactions are not included, the first two peaks are placed at $\sim$~3.17 and 3.18~eV, the values of the direct band gaps at $\Gamma$ and $V_x$. The scenario changes in the GW-BSE framework: two intense exciton peaks are present at 2.2 eV and 2.6 eV, below the indirect G$_0$W$_0$ gap of 2.85~eV.
These two excitons are still originated from critical points but the e-h interactions induce different intensity, and have different binding energies. The first exciton at 2.2~eV shows a very strong binding energy of $E_b$ = $1$~eV and is assigned to direct transitions at $\Gamma$. The second exciton peak (of lower intensity) is originated from the direct transitions between the valleys $V_x$ and shows a much weaker binding energy of $E_b$ = 0.5 eV. This result confirms the qualitative hydrogen-like picture in which the higher binding energies arises from the higher effective mass at $\Gamma$ with respect to $V_x$ (the numerical values in Tab.~\ref{tab-m}).
In contrast, in the armchair case, the top valence- bottom conduction band transitions at $V_x$ and $\Gamma$ points are very attenuated by the dipole coupling term (as displayed at GW-RPA level). The GW-BSE absorption spectrum calculations shows strong excitonic effects that lead single-particle transitions with energies higher than the band gap to the continuum of bottom of conduction band. Excitons living in the band gap are almost suppressed.

\begin{figure}[!htb]
    \centerline{
    \includegraphics[scale=0.30]{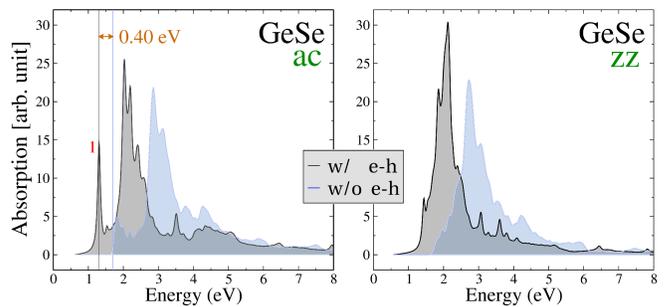}}
   \caption{\small (Color online) Absorption spectra of GeSe monolayer with (gray spectrum) and without (blue spectrum) electron-hole interactions for light polarized along zigzag (zz) and armchair (ac) in-plane directions. There is one excitonic state along $x$ (peak 1), due to a direct transition at the $V_x$ valleys (along the $\Gamma$-X direction).}
   \label{abs_GeSe}   
 \end{figure}

In GeSe monolayer, absorption spectrum calculations  (Fig.~\ref{abs_GeSe}) provide different outcomes. As far as armchair direction is concerned, in the absorption spectra calculated with the GW-RPA the first peak is at 1.70~eV, originated at $V_x$. This is well separated from the most intense peak at 3.14 eV, originating at the $\Gamma$ point. In GW-BSE, the only exciton in the fundamental band gap is placed at 1.30~eV, with a binding energy $E_b$ = $0.4$~eV. In the zig-zag direction, the excitonic effects determine a general red shift of the one particle excited states with an increase in intensity of the continuum of the transitions.

\begin{figure*}[!htb]
    \centerline{
    \includegraphics{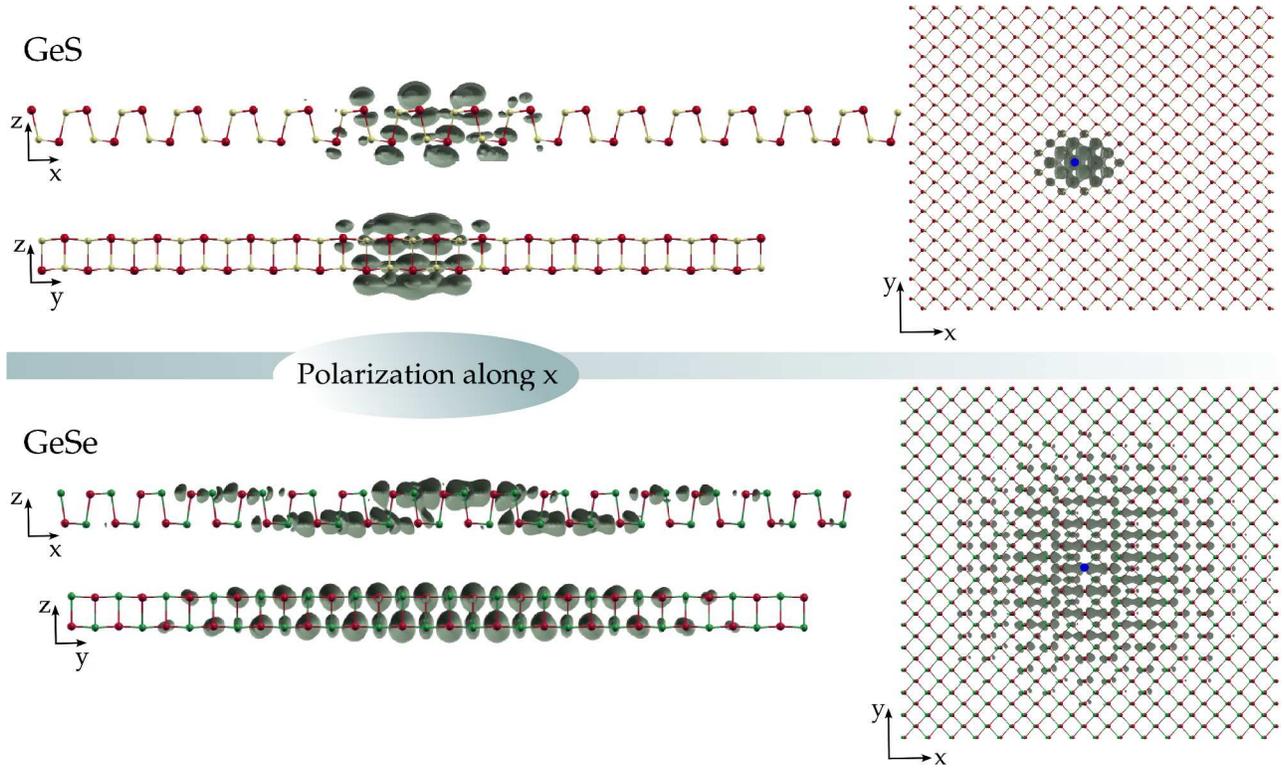}}
   \caption{\small (Color online) Normalized squared exciton wavefunction ($\AA^{-2}$) of excitons in GeS and GeSe for incident light polarized along the in-plane $x$ direction. The plot size corresponds to 16$\times$16 lattice spacings.}
   \label{excitons}
 \end{figure*}

In order to better understand the nature of the first peak excitons in armchair directions in GeS and GeSe monolayer, the normalized squared electron-hole wavefunctions $\Psi ({\bf r_e},{\bf r_h},)$ have been shown in Fig.~\ref{excitons}. The hole is placed in the center (blue spot). The plot sizes correspond to a 16$\times$16 unit cell. It is clear that the low energy exciton is more spatially localized in GeS than in GeSe, as should be expected by its higher $E_b$ and more ionic behaviour of sulfur atoms with respect to selenium atoms which reduce the electronic screening~\cite{PhysRevLett.110.016402}. 


\begin{table}[t]
\centering
\renewcommand{\arraystretch}{1.4}
\begin{tabular}{cccc}
\hline 
       &  \multicolumn{3}{c}{fundamental band Gap} \\ \hline
       &    DFT-GGA    & HSE      & G$_0$W$_0$     \\
GeS    &     1.70      & 2.45     &  2.85          \\
GeSe   &     1.14      & 1.79     &  1.70          \\ \hline 

\end{tabular}
\caption{\small Gap energies for DFT-GGA, HSE and G$_0$W$_0$ methods.}
\label{tab-gap}
\end{table}

\begin{table}[t]
\centering
\renewcommand{\arraystretch}{1.4}
\begin{tabular}{ccccccccccc}
\hline
    & & \multicolumn{4}{c}{$V_x$} &\phantom{a} &   \multicolumn{4}{c}{$\Gamma$} \\  \cline{3-6} \cline{8-11}
    
    &  & \multicolumn{2}{c}{m$_e^*$/m$_0$}   & \multicolumn{2}{c}{m$_h^*$/m$_0$} &    & \multicolumn{2}{c}{m$_e^*$/m$_0$} & \multicolumn{2}{c}{m$_h^*$/m$_0$}  \\ \cline{3-4} \cline{5-6} \cline{8-9} \cline{10-11}
    &  &    x  &  y       &   x  &   y       &    &   x   &  y       &  x   &   y      \\ 
GeS & &  0.27 & -0.50    & 0.23 & -0.46     &    &  0.57 & -1.99    & 0.65 & -1.39      \\
GeSe& &  0.20 & -0.22    & 0.17 & -0.20     &    &  1.28 & -2.75    & 2.83 & -4.17      \\ \hline

\end{tabular}
\caption{\small Effective masses of electrons (m$_e^*$/m$_0$) and holes (m$_h^*$/m$_0$) of valleys located at $\Gamma$ and at the $V_x$ valley (along the $\Gamma$-X direction), for the x and y in-plane directions.}
\label{tab-m}
\end{table}

{\it Mott-Wannier model--}
The GW-BSE framework used so far to investigate the optical properties of GeS and GeSe monolayers, provides information on excitonic properties of the isolated 2D systems. Nevertheless, in order to further clarify the nature of the excitons above described (if Frenkel or Mott-Wannier) and to evaluate the substrate effects in the binding energies, we have applied the model in Ref.~\onlinecite{PhysRevB.90.075429} for excitons in anisotropic 2D semiconducting crystals.

When dealing with 2D materials, one must exercise caution as the Coulomb term is replaced by the Keldysh-like potential due to the in-plane screening. The potential $\phi_{\rm 2D}$ felt by an electron in a 2D dielectric was obtained in Ref.~\onlinecite{Cudazzo2011dsi} for a 2D sheet in vaccum and extended in Ref.~\onlinecite{PhysRevB.90.075429} including the effects of a substrate bulk dielectric. For the later case, $\phi_{\rm 2D}$ has the form:

\begin{equation}
\phi_{\rm 2D}(r) = \frac{\pi q}{2\kappa r_0}\left[H_0\left(\frac{r}{r_0}\right) - Y_0\left(\frac{r}{r_0}\right)\right].
\label{phi2d}
\end{equation}
$H_0(r)$ and $Y_0(r)$ are Struve and Bessel functions, respectively, $r_0 = 2\pi\xi/\kappa$ is a length scale depending on the substrate dielectric constant $\epsilon$, $\kappa = (1+\epsilon)/2$.

From the asymptotic behaviour of $H_0(r)$ and $Y_0(r)$ it is determined that, at large $r$, the $\phi_{\rm 2D}$ interaction follows the usual $1/r$ form, while at small distances it diverges logarithmically.~\cite{Keldysh1979cii,Cudazzo2011dsi,Berkelbach2013ton,Castellanos_Gomez2014iac} Since the logarithmic well is more gradual than $1/r$, this results in substantially smaller binding energies. 

Once the correct potential is chosen, it is important to keep in mind that the reduction of  the two-body exciton problem to a single-body central potential problem is only applicable to Mott-Wannier excitons where the wave function is much larger than the lattice constant. The reason behind this requirement is the fact that the potential (\ref{phi2d}) treats the system as a continuous medium. With these considerations, it is possible to determine exciton binding energies of 2D anisotropic materials, with and without substrate effects, as detailed in Ref.~\onlinecite{PhysRevB.90.075429}. 

\begin{figure}[!htb]
    \centerline{
    \includegraphics[scale=0.25]{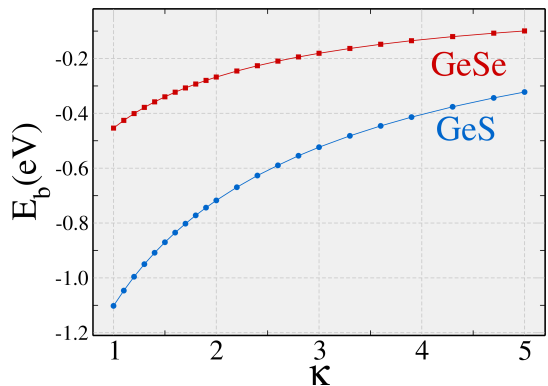}}
   \caption{\small (Color online) Exciton binding energy of GeS and GeSe monolayers as a function of the substrate dielectric constant $\kappa$=(1+$\epsilon$)/2.}
   \label{ebXkappa}
\end{figure}

The dependence of the exciton binding energies on the substrate dielectric constant $\kappa$, for single-layer GeS and GeSe, are presented in the plot in Fig.~\ref{ebXkappa}, for $\kappa$ ranging from 1 to 5. For the case of isolated layers, for which $\kappa$~=~1, the model gives $E_b$~=~1.10 and 0.45~eV for GeS and GeSe, respectively. This result is in fairly good agreement with the values obtained with our ab-initio GW-BSE calculations, confirming that these excitons are of Mott-Wannier character.

Screening effects introduced by the substrate decrease the binding energies with increasing $\kappa$. As an example, a reasonable choice is to consider the layers deposited on SiO$_2$ substrate ($\kappa \approx$~2.4). In this case, the exciton binding energies are reduced to 0.60 and 0.22~eV for GeS and GeSe, placing phosphorene in the middle-way of these two materials, with an exciton binding energy of 0.4~eV for the same substrate~\citep{PhysRevB.90.075429}.

{\it Conclusions--}\label{Conclusions}
Quasiparticle bandstructure and excitonic properties of orthorhombic two-dimensional GeS and GeSe are investigated by first-principles calculations. The $G_{0}W_{0}$ formalism indicates that 2D-GeS is a indirect gap material with energy gap of 2.85~eV, while 2D-GeSe is characterized by a minimum direct gap of 1.70~eV. 

However, the optical spectra of both materials is dominated by excitonic effects.
GeS $\Gamma$-point excitons have a remarkably large binding energy of 1 eV,
shifting the optical absorption threshold to 2.2 eV, in the visible range 
(rather than at 3.17 eV, in the near ultra-violet region, as expected from the quasiparticle gap).
Additionaly, the two gap excitions at $\Gamma$ and at the $V_x$ valley (along the $\Gamma$-X direction)
couple with light polarized along the $x$ direction.
Thus, between 2.2 and 2.6 eV, optical absorption is polarized.

For GeSe, the only exciton in the gap, with a binding energy of 0.4 eV, corresponds to 
the $V_x$ valley.
The binding energy of this exciton is more robust to external dielectric screening than in GeS.

Despite their strong binding, excitons binding energies are found to be in agreement with a 2D Mott-Wannier model\cite{PhysRevB.90.075429}.

\section*{Acknowledgements}
This work was supported by the National Research Foundation, Prime Minister Office, Singapore,
under its Medium Sized Centre Programme and CRP
award ``Novel 2D materials with tailored properties: beyond graphene" (Grant number R-144-000-295-281).
The first-principles calculations were carried out on the GRC high-performance computing facilities.


\end{document}